\documentstyle[12pt,epsf,epsfig]{article}
\def\beq{\begin{equation}}
\def\eeq{\end{equation}}
\def\bea{\begin{eqnarray}}
\def\eea{\end{eqnarray}}
\def\bq{\begin{quote}}
\def\eq{\end{quote}}

\def\gappeq{\mathrel{\rlap {\raise.5ex\hbox{$>$}}
{\lower.5ex\hbox{$\sim$}}}}

\def\lappeq{\mathrel{\rlap{\raise.5ex\hbox{$<$}}
{\lower.5ex\hbox{$\sim$}}}}

\begin{document}
\pagestyle{empty}
\begin{flushright}
{CERN-TH/2002-132}\\
{FISIST/11-2002/CFIF}\\
hep-ph/0206148\\
\end{flushright}
\vspace*{5mm}
\begin{center}
{\large {\bf Tau Flavour Violation in Sparticle Decays at the LHC}} \\
\vspace*{1cm}
{\bf D.~F.~Carvalho$^1$, J.~Ellis$^2$, M.~E.~G{\'o}mez$^1$, S.~Lola$^2$} 
and
{\bf J.~C.~Rom{\~a}o$^1$}\\
\vspace{0.3cm}
$^1$ CFIF, Departamento de Fisica, Instituto Superior T\'ecnico, \\
Av. Rovisco Pais, 1049-001~Lisboa, Portugal 

$^2$ CERN, CH-1211 Geneva 23, Switzerland \\
\vspace*{2cm}
{\bf ABSTRACT} \\ 
\end{center}
\vspace*{5mm}
\noindent

We consider sparticle decays that violate $\tau$ lepton number, motivated
by neutrino oscillation data. We work in the context of the constrained
minimal supersymmetric extension of the Standard Model (CMSSM), in which
the different sleptons have identical masses at the GUT scale, and
neutrino Dirac Yukawa couplings mix them. We find that the branching ratio
for decay of the heavier neutralino $\chi_2 \to \chi + \tau^\pm \mu^\mp$
is enhanced when the LSP mass $m_\chi \sim m_{\tilde \tau_1}$, including
the region of CMSSM parameter space where coannihilation keeps the relic
$\chi$ density within the range preferred by cosmology. Thus $\chi_2 \to
\chi + \tau^\pm \mu^\mp$ decay may provide a physics opportunity for
observing the violation of $\tau$ lepton number at the LHC that is
complementary to $\tau \to \mu + \gamma$ decay. Likewise, $\chi_2 \to \chi
+ e^\pm \mu^\mp$ decay is also enhanced in the coannihilation region,
providing a complement to $\mu \to e + \gamma$ decay. 

\vspace*{5cm}
\noindent

\begin{flushleft} CERN-TH/2002-132 \\
June 2002
\end{flushleft}
\vfill\eject

\setcounter{page}{1}
\pagestyle{plain}


\section{Introduction}

Mixing between different neutrino flavours has now been amply confirmed by
experiments on both atmospheric~\cite{skatm} and solar~\cite{sksol,sno}
neutrinos. The distorted zenith-angle distributions observed by
Super-Kamiokande provided a `smoking gun' for atmospheric-neutrino
oscillations, establishing that they are most likely due to near-maximal
$\nu_\mu - \nu_\tau$ mixing. More recently, SNO has provided two `smoking
guns' for solar-neutrino oscillations, providing direct evidence for
near-maximal $\nu_e \to \nu_{\mu, \tau}$ oscillations~\cite{snofit}
through its measurements of the charged- and neutral-current scattering
rates.

These observations lead one to expect the corresponding charged-lepton
numbers to be violated at some level. However, the rates for such
processes would be unobservably small if neutrino masses were generated by
the seesaw mechanism~\cite{seesaw} and there was no lower-energy physics
beyond the Standard Model. However, the naturalness of the gauge
hierarchy, grand unification of the gauge couplings and the relic density
of supersymmetric dark matter all suggest that supersymmetry should appear
at an energy scale $\lappeq 1$~TeV. This suggests that processes violating
the different charged lepton numbers might be observable in low-energy
experiments. Indeed, charged-lepton-number violating processes could occur
at embarrassingly large rates if the soft supersymmetry-breaking masses of
the squarks and sleptons were not universal. For this reason, it is often
assumed that these masses are equal at the grand-unification scale, as in
the constrained minimal supersymmetric extension of the Standard Model
(CMSSM).

Even in this case, renormalization of the soft supersymmetry-breaking
slepton masses would occur in the minimal supersymmetric version of the
seesaw model for neutrino masses, thanks to the Dirac Yukawa couplings of
the neutrinos~\cite{bm}. These are active in the renormalization-group
equations at scales between the GUT scale and the heavy singlet-neutrino
mass scale, and are not expected to be diagonal in the same basis where
the light leptons are flavour-diagonal. This scenario provides the minimal
credible amount of charged-lepton-flavour violation:  it could be enhanced
by GUT interactions and/or non-universal slepton masses at the GUT scale.

Many signatures for charged-lepton-flavour violation have been considered
in this scenario~\cite{review,h1,ci,nlfv,CEGL,taumug}, including $\mu \to
e \gamma$ and related decays, $\tau \to \mu \gamma$ and $\tau \to e
\gamma$ decays. Any or all of these may be favoured by the (near-)maximal
mixing observed amongst the corresponding neutrino species.  Other things
being equal, one expects these decays to be relatively large when the soft
supersymmetry-breaking masses $m_{1/2}$ and/or $m_0$ are relatively small,
as has been borne out in specific model-dependent studies. Another
possibility that has been considered is the decay $\chi_2\to \chi + e^\pm
\mu^\mp$ \cite{sleptonoscemu,Hisano}, where $\chi$ is the lightest
neutralino, assumed here to be the lightest supersymmetric particle (LSP),
and $\chi_2$ is the second-lightest neutralino. It has been argued that
this decay might have a rate observable at the LHC for certain choices of
the CMSSM parameters.

In this paper, we consider the alternative decay $\chi_2 \to \chi +
\tau^\pm \mu^\mp$ \cite{sleptonoscmutau}. This has certain theoretical
advantages over the decay $\chi_2 \to \chi + e^\pm \mu^\mp$ considered
previously, since the feedthrough into the charged-lepton sector may be
enhanced by larger Dirac Yukawa couplings and/or lighter singlet-neutrino
masses, as compared to the $\nu_\mu - \nu_e$ sector, if neutrino masses
exhibit the expected hierarchical pattern, and $\nu_\tau - \nu_\mu$ mixing
is also known to be essentially maximal. On the other hand, the decay
$\chi_2 \to \chi + \tau^\pm \mu^\mp$ has a less distinctive experimental
signature than $\chi_2 \to \chi + e^\pm \mu^\mp$. Both decays should be
explored at the LHC and a possible linear $e^+ e^-$ collider, and which
mode offers better prospects may depend on the neutrino-mass model and the
experiment.

We find that the branching ratio for $\chi_2 \to \chi + \tau^\pm \mu^\mp$
decay is enhanced when $m_{\chi_2} > m_{\tilde \tau_1} > m_\chi$, where
${\tilde \tau_1}$ is the lighter stau slepton. This occurs in a wedge of
the $(m_{1/2}, m_0)$ parameter plane in the CMSSM that is complementary to
that explored by $\tau \to \mu \gamma$.  The region of CMSSM parameter
space where this enhancement occurs includes the region where $\chi -
{\tilde \ell}$ coannihilation suppresses the relic density $\Omega_\chi$,
keeping it within the range $0.1 < \Omega_\chi h^2 < 0.3$ preferred by
astrophysics and cosmology, even if $m_{1/2}$ is comparatively large. The
interest of this coannihilation region has been accentuated by the latest
experimental constraints on the CMSSM, such as $m_h$ and $b \to s \gamma$
decay, which disfavour low values of $m_{1/2}$.  We show that the
branching ratio for $\chi_2 \to \chi + \tau^\pm \mu^\mp$ decay may be a
large fraction of that for the flavour-conserving decay $\chi_2 \to \chi +
\mu^\pm \mu^\mp$. An analogous enhancement is expected for the
flavour-violating decay $\chi_2 \to \chi + e^\pm \mu^\mp$ considered by
other authors~\cite{sleptonoscemu,Hisano}, although the absolute branching
ratio is expected to be smaller. Nevertheless, this decay may provide
another way of probing lepton flavour violation in the coannihilation
region.

\section{Calculational Framework}

We assume the minimal supersymmetric extension of the seesaw mechanism for 
generating neutrino masses, in which there are three heavy 
singlet-neutrino states $N_i$, and the leptonic sector of the 
superpotential is:
\begin{equation}
W = N^{c}_i (Y_\nu)_{ij} L_j H_2
  -  E^{c}_i (Y_e)_{ij}  L_j H_1
  + \frac{1}{2}{N^c}_i {\cal M}_{ij} N^c_j + \mu H_2 H_1 \,,
\label{leptonW}
\end{equation}
where $Y_\nu$ is the neutrino Dirac Yukawa coupling matrix, ${\cal 
M}_{ij}$ is the Majorana mass matrix for the $N_i$, the $L_j$ and $H_I$ 
are lepton and Higgs doublets, and the $E^{c}_i$ are singlet 
charged-lepton supermultiplets. 
The superpotential of  the effective 
low-energy theory, obtained after the decoupling of 
heavy neutrinos is  \cite{valle}
\begin{eqnarray}
\label{weff}
W_{eff}&=& L_i H_2 \left(Y_\nu^T \left({\cal M}^{D}\right)^{-1} 
Y_\nu\right)_{ij} L_j H_2 
  -  E^{c}_i (Y_e)_{ij}  L_j H_1.
\label{LEET}
\end{eqnarray}
In the basis where the charged leptons and
the heavy neutrino mass matrices are diagonal,
\bea 
{\cal M}_\nu={Y}_\nu^T \left({\cal M}^{D}\right)^{-1} {
Y}_\nu v^2 \sin^2\beta
\label{seesaw1}
\eea
where the $v=174$ GeV and  $\tan\beta=v_2/v_1$.

As mentioned above, we work in the context of the CMSSM, where the soft 
supersymmetry-breaking masses of the charged and neutral sleptons are 
assumed to be universal at the GUT scale, with a common value $m_0$. In 
the leading-logarithmic approximation, the non-universal renormalization 
of the soft supersymmetry-breaking scalar masses is by an amount
\begin{equation}
\left(\delta m_{\tilde L}^2\right)_{ij} \approx 
-\frac{1}{8\pi^2}(3m_0^2 + A_0^2)
({ Y_\nu^\dagger}{ Y_\nu})_{ij}
\log\frac{M_{GUT}}{M_{N_i}}.
\label{nonuniv}
\end{equation}
We note that, in this approach, non-universality in the soft 
supersymmetry-breaking left-slepton masses is much larger than that in 
right-slepton masses when the trilinear soft supersymmetry-breaking
parameter $A_0 = 0$, as we assume here~\footnote{In the case $A_0 \ne 0$, 
this parameter would also be renormalized analogously to $m_{\tilde L}^2$ 
(\ref{nonuniv}).}. The pattern of charged-lepton-flavour violation 
induced by renormalization depends on the details of $(Y_\nu)_{ij}$.

In plausible mixing textures,
the renormalization of the soft supersymmetry-breaking parameters at low 
energies can be understood approximately in terms of the dominant 
non-universality in the third-generation left-slepton mass:
\begin{equation}
m^2_{0_{LL}} =  {\rm diag} (m_0^2, m_0^2, x \times m_0^2),
\label{nonuniv2}
\end{equation}
where a typical value of the non-universality factor is $x \sim 0.9$. 
Correspondingly, we assume there is an off-diagonal ${\tilde \tau_L} - 
{\tilde 
\mu_L}$ mixing term in the soft mass-squared matrix:
\begin{equation}
\Delta m^2_{0_{LL}} = (1-x) m^2_0 {\sin (2\phi) \over 2},
\label{offdiag}
\end{equation}
where $\phi$ is the mixing angle between the second and third generation
in the charged-lepton Yukawa matrix. For the type of non-universalities
introduced in (\ref{nonuniv2}), this angle can be quite large
without entering in conflict with the current bounds for $\tau\rightarrow
\mu \gamma$, though in this case large mixing in the 2-3 sector must be
combined with a small mixing angle between the first and second
generation, due to the very restrictive bound in the $\mu\rightarrow e
\gamma$ decay \cite{mix-magn}. This mixing leads to lepton-flavour
violation $\sim \sin^2 (2\phi)$, as long as $\sin (2\phi)$ is not too
large~\footnote{We have checked that this parametrization is generally a
very good approximation for $\tau \to \mu \gamma$ decay, as well as for
$\chi_2 \to \chi \tau^\pm \mu^\mp$.}.

We give below numerical results for sample choices of the parameters $(x,
\phi)$ that may be representative of the possibilities in specific models.
We also show how the results vary as $(x, \phi)$ are varied.

In the following, we consider mixing between the left-handed $\tau$- and
$\mu$-flavoured sleptons, but ${\tilde \tau} - {\tilde e}$ mixing might
also be present, or even favoured in some models. In such cases, the
results would be rather similar to those we present, simply with $\mu$
replaced by $e$ in the $L_\tau$-violating decay modes studied.

We consider the following flavour-violating and -conserving $\chi_2$ 
decays:
\begin{eqnarray}
\chi_2 & \rightarrow  \tilde{\ell}_i \ell_j  
 \rightarrow \chi \ell_i^{+} \ell_j^{-}  \; , \;
\chi_2 & \rightarrow  \tilde{\nu}_i \nu_j
 \rightarrow \chi \nu_i \bar{\nu}_j 
\\
\chi_2 & \rightarrow    \chi Z  \rightarrow  
\chi \ell^{+}_i \ell^{-}_i  \; , \;
\chi_2 & \rightarrow    \chi Z  \rightarrow  \chi \nu_i  \bar{\nu}_i
\end{eqnarray}
\begin{equation}
\chi_2 \rightarrow \chi h  \rightarrow \chi \ell^{+}_i \ell^{-}_i  
\label{3h}
\end{equation}
The first two decays are the only ones in which  flavour 
violation may be expected, and it would of course be unobservable in 
$\chi_2 \rightarrow \chi \nu \bar{\nu}$ decay. The intermediate sleptons 
are produced on-shell if they are 
lighter than the $\chi_2$, while the $Z$ and the $h$ are always on-shell 
for the range of parameters that are of interest to us. 
Slepton exchanges and $h$ decays may give significantly different rates
for the various flavor-conserving decays $\chi_2 \rightarrow \chi
\ell^{+}_i
\ell^{-}_i$, suppressing the cases $\ell = \mu, e$ relative to the
case $\ell = \tau$, an effect we see in subsequent plots.

Our calculations are similar to \cite{Hisano}, except that we also include
the Yukawa interactions, which are relevant for decays into $\tau$ leptons
at large $\tan\beta$. Furthermore, we include finite-width effects in our
calculations of $Z^0$ and slepton exchanges. The neutralino and slepton
widths, which arise mainly from two-body decays, were calculated using the
{\tt ISAJET} package~\cite{ISA}, and a check with {\tt
calcHEP}~\cite{calchep} found good agreement. For the decays $\chi_2 \to
\chi + \mu^\pm \mu^\mp$, we found good agreement between our code and {\tt
calcHEP}, once we incorporated the {\tt VEGAS} adaptive Monte Carlo
programme for the momentum integrals in three-body decays. The results
from {\tt VEGAS} differ by several orders of magnitude from those obtained
using {\tt ISAJET} for $\chi_2 \to \chi + \mu^\pm \mu^\mp$ decay close to
the ${\tilde \tau}$ resonances.

For the decay $\chi_2 \to \chi + \tau^\pm \tau^\mp$ the channels mediated
by Higgses are important in the areas where $m_{\tilde{\chi}_2} -
m_{\tilde{\chi}} < m_{\tilde{\tau}_1} $. The widths have been obtained
using {\tt calcHEP}, after adding to the package the one-loop QCD
corrected Higgs widths from {\tt HDECAY}~\cite{dks}.  For
flavour-violating decays, our calculation agrees with {\tt calcHEP}, once
we modify the MSSM Lagrangian included in this package to allow
flavour-mixing among $\tilde{\tau}_1$, $\tilde{\tau}_2 $ and
$\tilde{\mu}_L $.

\section{Numerical Results}

\begin{figure}
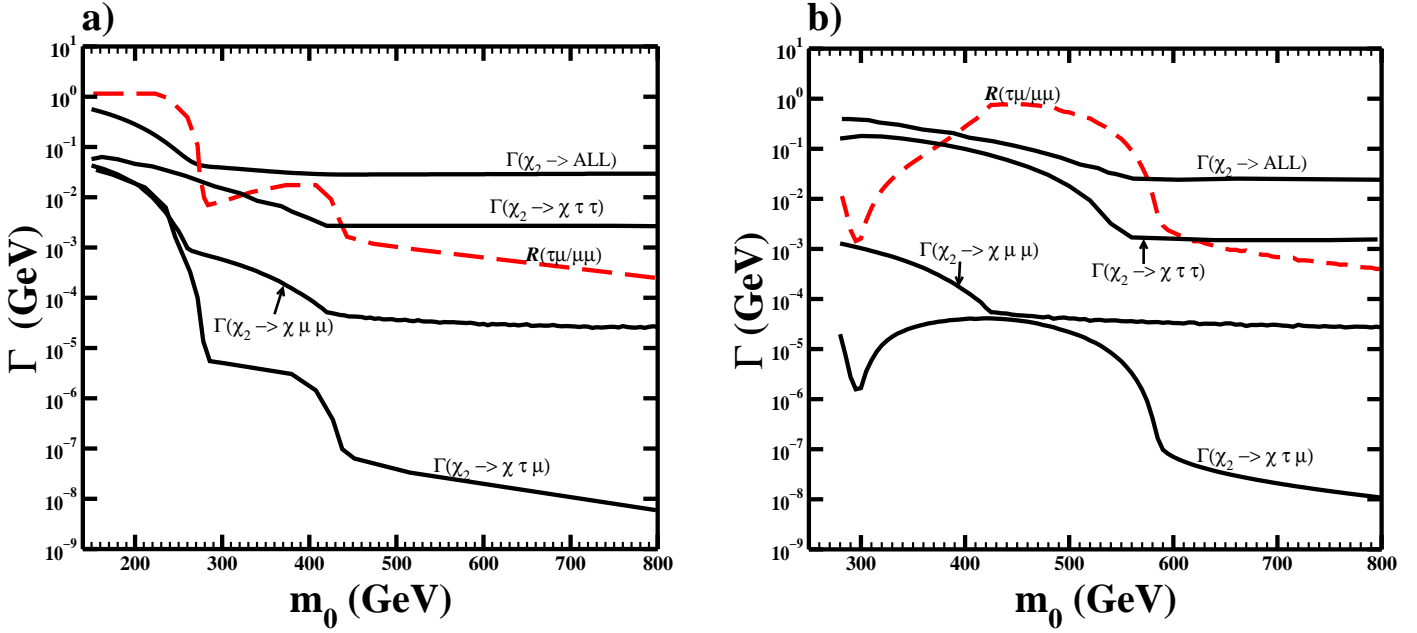

\hspace*{-.35in}
\begin{minipage}{8in}
\epsfig{file=width10p.eps,height=3.3in}
\hspace*{0.5 cm}
\epsfig{file=width40p.eps,height=3.3in} 
\hfill
\end{minipage} 
\hfill
\caption[]{\it
Comparison of flavour-changing and -conserving $\chi_2$ decay modes as
functions of $m_0$ for (a) $\tan \beta = 10, \mu > 0, m_{1/2} = 600$ GeV and 
(b) $\tan \beta = 40, \mu > 0, m_{1/2} = 600$ GeV. We assume for illustration 
a non-universality factor $x = 0.9$ and a mixing angle $\phi = \frac{\pi}{6}$.  
}
\label{fig:widths}
\end{figure}

The solid (black) lines in Fig.~\ref{fig:widths} denote the total
$\chi_2$ decay width, as well as the partial widths for the
flavour-violating and flavour-conserving decays, for the particular cases
(a) $\tan \beta = 10, \mu > 0, m_{1/2} = 600$~GeV and (b) $\tan \beta =
40, \mu > 0, m_{1/2} = 600$~GeV. In both plots, we make the
representative choices $x = 0.9$ and $\phi = \pi / 6$.  In
Fig.~\ref{fig:widths}(a), we see a first edge in the flavour-violating
width $\Gamma(\chi_2 \to \chi + \tau^\pm \mu^\mp)$ at $m_0 \sim 280$~GeV,
which is less pronounced in $\Gamma(\chi_2 \to \chi + \mu^\pm \mu^\mp)$
and almost absent in $\Gamma(\chi_2 \to \chi + \tau^\pm \tau^\mp)$.  
This reflects the dominant role of ${\tilde \tau_2} \sim {\tilde \tau_L}$
exchange in the flavour-violating case.  We also note a second edge when
$m_{\tilde \tau_1} = m_{\chi_2}$ at $m_0 \sim 430$~GeV, which is visible
in all the flavour-violating and flavour-conserving decays to $\chi$ and
leptons. The differences between $\Gamma(\chi_2 \to \chi + \tau^\pm
\tau^\mp)$ and $\Gamma(\chi_2 \to \chi + \mu^\pm \mu^\mp)$ are due, at
smaller $m_0$, to the different masses and couplings of the ${\tilde
\tau}_{1,2}$ and ${\tilde \mu}_{L,R}$ being exchanged, whilst the
differences at larger $m_0$ are due to $\chi_2 \rightarrow \chi + h$
decay.

We see in panel (b) of Fig.~\ref{fig:widths} features at $m_0 = 300$~GeV,
420~GeV and 580~GeV, corresponding to $m_{\chi_2} = m_{\tilde \tau_1},
m_{\tilde \mu_R}$ and $m_{\tilde \tau_2}$, respectively. The lowest and highest
features show up in $\Gamma(\chi_2 \to \chi + \tau^\pm \mu^\mp)$ and
$\Gamma(\chi_2 \to \chi + \tau^\pm \tau^\mp)$ and the middle feature in
$\Gamma(\chi_2 \to \chi + \mu^\pm \mu^\mp)$, as one would expect. We note
that $\Gamma(\chi_2 \to \chi + \tau^\pm \tau^\mp)$ may become relatively
large for $300~{\rm GeV} < m_0 < 580$~GeV, becoming the dominant $\chi_2$
decay mode. 

The analogous plot for $\tan \beta = 10, \mu < 0, m_0 = 600$~GeV is quite
similar to panel (a) of Fig.~\ref{fig:widths}, whilst that for $\tan
\beta = 30, \mu > 0, m_0 = 600$~GeV is intermediate between panels (a)
and (b). Hence these are representative of the possibilities for
flavour-violating $\chi_2$ decays.

The ratio of branching ratios $R(\tau\mu / \mu \mu) \equiv \Gamma(\chi_2
\to \chi + \tau^\pm \mu^\mp) / \Gamma(\chi_2 \to \chi + \mu^\pm \mu^\mp)$
is shown as (red) dashed lines in Fig.~\ref{fig:widths}(a,b). In panel
(a), the quantity $R(\tau\mu / \mu \mu)$ also exhibits clearly the first
edge at $m_0 \sim 280$~GeV. The second edge at $m_0 \sim 430$~GeV also
appears strongly, reflecting the facts that flavour violation appears
mainly in the left-slepton sector, and that the ${\tilde \tau_2}$ is
mainly ${\tilde \tau_L}$. We see that, for our choices of $x$ and $\phi$,
$R(\tau \mu / \mu \mu)$ may be of order unity for $m_0 < 270$~GeV, and $
\sim 10^{-2}$ for $m_0 < 430$~GeV. Only at larger $m_0$, where the $\chi_2
\rightarrow \chi + {\tilde \tau}$ decay becomes kinematically
inaccessible, does $R(\chi_2 \to \chi + \tau^\pm \mu^\mp)$ drop below
$10^{-3}$. In panel (b) of Fig.~\ref{fig:widths}, we see that $R(\tau \mu
/ \mu \mu) \sim 0.1$ to unity for $350~{\rm GeV} < m_0 < 580$~GeV,
dropping below $10^{-3}$ only for $m_0 > 600$~GeV.

\begin{figure}[h]
\begin{center}
\epsfig{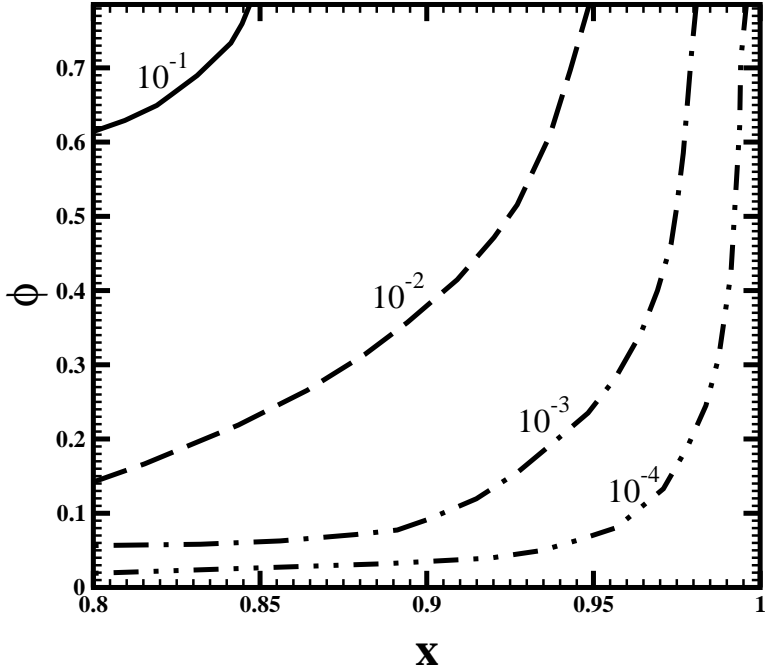}
\end{center}
\caption[]{\it
Contours of the ratio $R(\tau \mu / \mu \mu)$ of the branching ratios for 
the flavour-violating and flavour-conserving decays in the $x,~\phi$ 
plane, for
$\tan\beta = 30$, $\mu > 0$, $m_{1/2} = 400$  GeV and  $m_{0} = 200$  
GeV.
}
\label{fig:xphi}
\end{figure}

In Fig.~\ref{fig:xphi} we display contours of the ratio $R(\tau \mu
/ \mu \mu)$ of the branching ratios for the flavour-violating decay
$\chi_2 \to \chi + \tau^\pm \mu^\mp$ and the flavour-conserving decay
$\chi_2 \to \chi + \mu^\pm \mu^\mp$ in the $x, \phi$ plane, for the
particular choices $\tan\beta = 30$, $m_{1/2} = 400$~GeV and $m_{0} =
200$~GeV of the CMSSM parameters. We see that the previous choice $x =
0.9, \phi = \pi / 6$ is not particularly exceptional. To quite a good
approximation, $R(\tau \mu / \mu \mu)$ scales by the square of the factor
$(1-x) \sin (2\phi)$ shown in (\ref{offdiag}). This makes it relatively
easy to reinterpret our illustrative results in the context of any
specific flavour texture model that makes definite predictions for
$x$ and $\phi$.

\begin{figure}
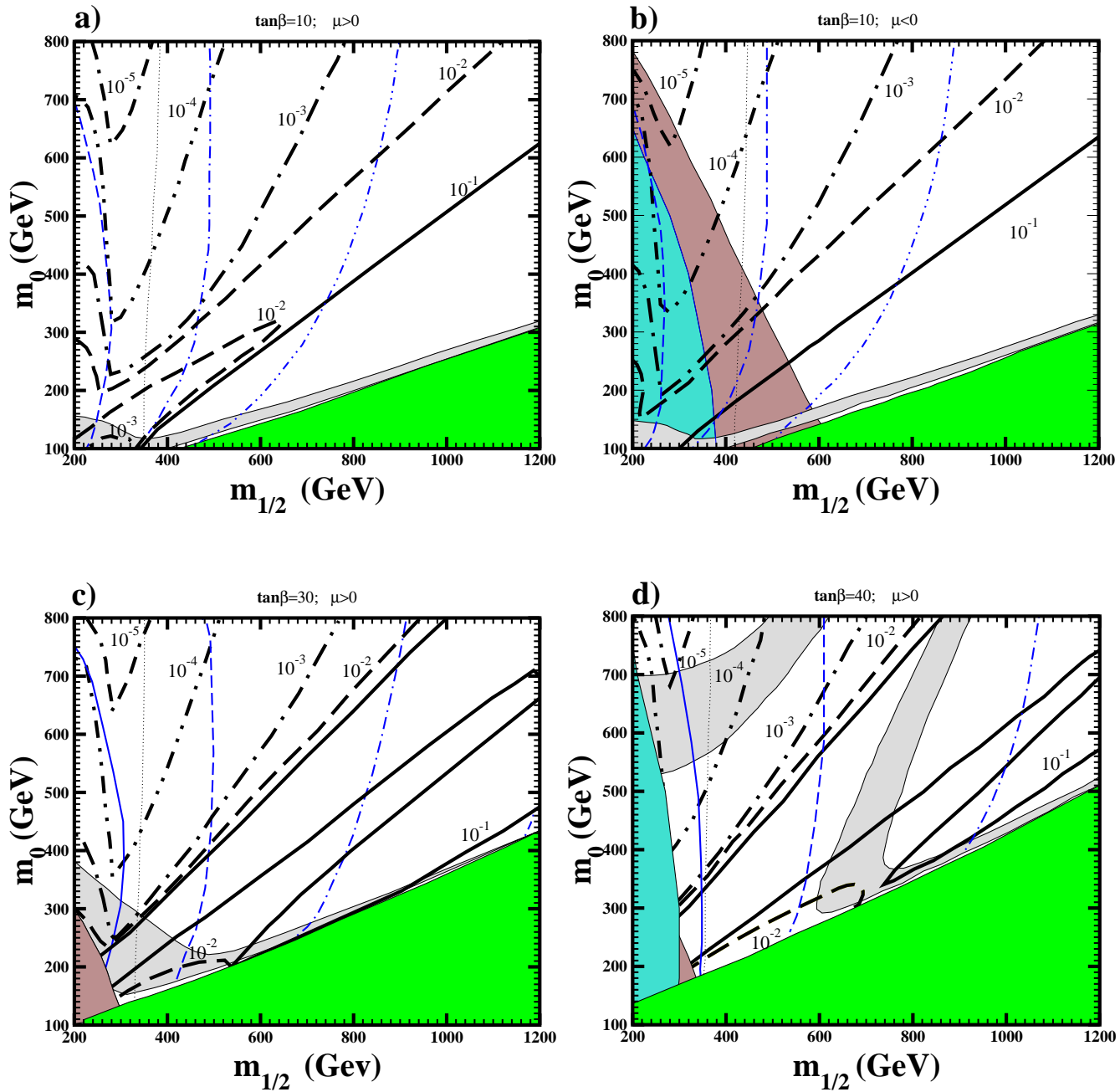

\vspace*{-0.75in}
\hspace*{-.35in}
\begin{minipage}{8in}
\epsfig{file=area10.eps,height=3.2in}
\epsfig{file=area10n.eps,height=3.2in} 
\hfill
\end{minipage}
\hspace*{-.35in}
\begin{minipage}{8in}
\vspace*{1.0 cm}
\epsfig{file=area30.eps,height=3.2in}
\epsfig{file=area40.eps,height=3.2in} \hfill
\end{minipage}
\vspace*{0.5 cm}
\caption{\label{fig:contours}
{\it 
Contours of $R(\tau \mu / \mu \mu)$ (thick black lines) and
$BR(\tau \to \mu \gamma)$ (thin blue lines) in the $(m_{1/2}, m_0)$ planes 
for
(a) $\tan \beta = 10, \mu > 0$, b) $\tan \beta = 10, \mu < 0$, (c) $\tan 
\beta = 30, \mu > 0$, (d)  $\tan \beta = 40, \mu > 0$,
for $x = 0.9$ and $\phi = \pi / 6$. 
The regions 
disallowed at low $m_{1/2}$ and $m_0$ by
the E821 measurement of $a_\mu$ at the 2-$\sigma$ level~\cite{BNLE821} are
dark (brown) shaded, the dark (green) shaded regions at large $m_{1/2}$ 
and low $m_0$ are 
excluded because
the LSP is the charged ${\tilde \tau}_1$, the light grey shaded
regions are those with $0.1 < \Omega_\chi h^2 < 0.3$ that are
preferred by cosmology (calculated using {\tt 
MICROMEGAS}~\cite{MICROMEGAS}), and the 
medium (blue)
shaded regions are
excluded by $b \to s \gamma$ \cite{jo2} 
and the dotted line is $m_h = 114.1$~GeV 
(calculated using {\tt FeynHiggs}~\cite{FeynHiggs}).  
}}
\end{figure}

We display in Fig.~\ref{fig:contours} contours of the branching ratio for
the flavour-violating decay $\tau \to \mu \gamma$ (thin blue lines) and
the flavour-violating ratio $R(\tau \mu / \mu \mu)$
(thick black lines) in the $(m_{1/2}, m_0)$ planes for different choices
of $\tan \beta$ and the sign of $\mu$.  In each case, we have again made
the representative choices $x = 0.9$ and $\phi = \pi / 6$. 

The contours where $R(\tau \mu / \mu \mu) = 10^{-1}, 10^{-2}, 10^{-3},
10^{-4}$ and $10^{-5}$ are shown as thick black solid, dashed, dot-dashed,
dot-dot-dashed and dot-dashed-dashed lines.  We also display contours of
$BR(\tau \to \mu \gamma) = 10^{-6}, 10^{-7}, 10^{-8}, 10^{-9}$ as thin
blue solid, dashed, dot-dashed and dot-dot-dashed lines. We see that large
${\tilde \mu} - {\tilde \tau}$ (or ${\tilde e} - {\tilde \tau}$) mixing is
not excluded by the present upper limits on $BR(\tau \to \mu (e) \gamma)$,
which are both just above $10^{-6}$. We also recall that the $\chi_2$ is
observable at the LHC in cascade decays of heavier
sparticles~\cite{HPetal} for many choices of CMSSM
parameters~\cite{Bench}. We see immediately from Fig.~\ref{fig:contours}
that the regions where $\chi_2 \to \chi + \tau^\pm \mu^\mp$ may be
observable at the LHC (or a future linear $e^+ e^-$ collider?), perhaps
where $R (\tau \mu / \mu \mu) \gappeq 10^{-2}$, are largely complementary
to those where $\tau \to \mu \gamma$ may be observable at the LHC or a $B$
factory, perhaps where $BR(\tau \to \mu \gamma) \gappeq 10^{-8}$.

The darker (green) shaded regions in the bottom right corners of the
panels in Fig.~\ref{fig:contours} are excluded because there the LSP is
the lighter stau: ${\tilde \tau_1}$. Such a charged LSP would be in
conflict with basic astrophysics. The lighter (grey) shaded regions are
those in which the cosmological relic density of the neutralino LSP $\chi$
is in the range preferred by cosmology: $0.1 < \Omega_\chi h^2 < 0.3$ as
calculated using {\tt MICROMEGAS}~\cite{MICROMEGAS}, and in agreement with
our previous calculations~\cite{jo1,jo2,mlaz}, where $h \sim 0.7$ is the
current Hubble expansion rate in units of 100~km/s/Mpc. In each panel,
there is a region at small $m_{1/2}$ that is disfavoured by laboratory
experiments.  The regions at small $m_{1/2}$ excluded by the $b \to s
\gamma$ decay rate are medium (blue) shaded, the regions disfavoured by
$g_\mu - 2$ at small $m_{1/2}$ and $m_0$ are darker (brown) shaded, and
the (dotted) line is where $m_h = 114.1$~GeV as calculated using {\tt
FeynHiggs}~\cite{FeynHiggs}. Together, these constraints favour the
coannihilation strip where $m_{\tilde \tau_1} \sim m_\chi$ in all the
panels, and the channels at large $m_{1/2}$ and $m_0$ in panel (d) where
direct-channel $\chi \chi \to A,H$ annihilation is relatively rapid.

In panel (a), for $\tan \beta = 10, \mu > 0$, the LEP search for the Higgs
boson disfavours $m_{1/2} \lappeq 360$~GeV. In panel (b), for $\tan \beta
= 10, \mu < 0$, the observed rate for $b \to s \gamma$ decay excludes
$m_{1/2} \lappeq 300$~GeV, the LEP search for the Higgs boson disfavours
$m_{1/2} \lappeq 430$~GeV, and $g_\mu - 2$ excludes a triangle extending
up to $m_{1/2} \sim 600$~GeV. In panel (c)  for $\tan \beta = 30, \mu >
0$, the LEP Higgs limit disfavours $m_{1/2} \lappeq 340$~GeV, and the
other constraints are weaker. A similar pattern is repeated in panel (d),
for $\tan \beta = 40, \mu > 0$.

In cases (a, b, c), the only region of the $(m_{1/2}, m_0)$ plane that
survives these constraints is the strip parallel to the boundary of the
disallowed region, where $m_\chi / m_{\tilde \tau_1} \sim 1.1 - 1.2$, and
coannihilation keeps $\Omega_\chi h^2$ within the range allowed by
astrophysics and cosmology. This is precisely the region where $R(\tau
\mu/ \mu \mu)$ is maximized, and hence the chances of observing the decay
may be maximized. We do note, however, that $R(\tau \mu / \mu \mu)$ has a
tendency to fall as $m_{1/2}$ increases along this strip, which is
apparent in panels (c) and (d). We further note in panel (d) that $R(\tau
\mu / \mu \mu) \gappeq 10^{-2}$ also on the right side of the rapid $\chi
\chi \to A, H$ annihilation channel, but may be significantly lower on the
left side of this channel.

\section{Conclusions}

We have demonstrated in this paper that the decay $\chi_2 \to \chi
\tau^\pm \mu^\mp$ provides an opportunity to look for $\tau$ flavour
violation at the LHC that is largely complementary to the search for
$\tau \to \mu \gamma$. Essentially all the above analysis would apply
also if the slepton mixing texture favours $\chi_2 \to \chi \tau^\pm
e^\mp$ and $\tau \to e \gamma$ over $\chi_2 \to \chi \tau^\pm \mu^\mp$ and
$\tau \to \mu \gamma$: it is even possible that both $\chi_2 \to \chi
\tau^\pm \mu^\mp/e^\mp$ decays may be observable at the LHC.

We have phrased this analysis as model-independently as possible. 
Specific models will predict values for the mixing parameters $x$ and 
$\phi$, and the scaling of our results with these parameters is quite 
simple. We would expect the relevant mixing parameters to be much smaller 
in the case of $\chi_2 \to \chi \mu^\pm e^\mp$ decay, but the 
corresponding $R(\mu e /\mu \mu)$ would be enhanced in a similar region 
of the CMSSM parameter space.

We note that the $\mu^\mp$ produced in $\chi_2 \to \chi \tau^\pm \mu^\mp$
decay are likely to have significant transverse momentum, and any event
in which the $\chi_2$ is produced is likely to have considerable missing
transverse energy and jet activity associated with the decays of other
sparticles. Therefore, we do not expect such events to be suppressed
badly at the trigger level at the LHC, though it might be more difficult
to see $\chi_2 \to \chi \tau^\pm e^\mp$ decays. However, a detailed
simulation goes beyond the scope of this paper. There should be even less
problem seeing $\chi_2 \to \chi + \tau^\pm \mu/e^\mp$ decays at a linear
$e^+ e^-$ collider. We therefore urge more detailed simulations of this
decay mode for this machine, as well as for the LHC.

\end{document}